\documentstyle[12pt,moriond]{article}
\input{epsf}
\begin{document}
\heading{DENSITY PROFILES OF GALACTIC DARK HALOS}

\author{Ewa L. \L okas} {Nicolaus Copernicus Astronomical Center, Warsaw,
Poland} {\hspace{1cm}}

\begin{moriondabstract}
I propose a modification of the spherical infall model for the
evolution of density fluctuations with initially Gaussian probability
distribution and scale-free power spectra. I introduce a generalized form
of the initial density distribution around an overdense region and cut it
off at half the inter-peak separation accounting in this way for the
presence of neighbouring fluctuations. Contrary to the original
predictions of the model, resulting density profiles within virial radii
no longer have power-law dependence on the distance, but are well fitted
by the universal formula of changing slope obtained as a result of
$N$-body simulations. The profiles of halos of a given mass are in general
flatter than the corresponding ones from the simulations, but the trend of
steeper profiles for smaller masses is reproduced. The profiles of galaxy
size objects are in general in better agreement with N-body simulations
than those of larger ones.

\end{moriondabstract}

\section{Introduction}
High-resolution $N$-body simulations with power-law initial power spectra
\cite{nf} suggest that density profiles of dark halos in large range
of masses are well fitted by a simple universal formula
\begin{equation}    \label{p20}
    \frac{\rho(x)}{\rho_{\rm crit,0}} =
    \frac{\delta_{\rm char}}{(r/r_{\rm s})(1+r/r_{\rm s})^{2}}
\end{equation}
where $\rho_{\rm crit,0}$ is the present critical density of the
Universe and $\delta_{\rm char}$ is the characteristic density
\begin{equation}    \label{p22}
    \delta_{\rm char} = \frac{v c^3}{3 [{\rm ln} (1+c) - c/(1+c)]} ,
\end{equation}
with $v=200$.
The scale radius $r_{\rm s}$ is related to the virial radius $r_{v}$ (the
distance from the center of the halo within which the mean density is $v$
times the critical density) by $r_{\rm s} = r_{v}/c$ and $c$ is
the concentration, the only fitting parameter in the formula.

The density profile was observed to steepen from $r^{-1}$ near
the center of the halo to $r^{-3}$ at large distances. This result seemed
to contradict the prediction of the analytical spherical infall model
\cite{hs} (hereafter SIM) which for $\Omega=1$ (the only case considered
here) finds the profiles to be power-laws of the form $r^{-3(n+3)/(n+4)}$
where $n$ is the index of the initial power spectrum of density
fluctuations.

\section{The modified spherical infall model}
I will argue here that the discrepancy between the two approaches is
mainly due to the oversimplifications applied in the SIM. While such
assumptions as spherical symmetry of the initial density distribution and
the absence of peculiar velocities will be kept, the shape of the initial
density distribution can in fact be made more realistic. This
distribution is usually described by the expected overdensity within
$r_{\rm i}$ provided there is a peak (overdense region) of height $a
\sigma$ at $r_{\rm i}=0$, where $\sigma$ is the rms fluctuation of the
linear density field smoothed on scale $R$. If the initial probability
distribution of fluctuations is Gaussian and the filter is Gaussian the
general form of this quantity as a function of $x_{\rm i} = r_{\rm i}/R$
can be found
\begin{equation}
    \langle \Delta_{\rm i}(x_{\rm i})  \rangle = \frac{6 a \sigma}{(n+1)
    x_{\rm i}^{2}}  \left[ \
    _{1}F_{1} \left( \frac{n+1}{2}, \frac{3}{2}, - \frac{x_{\rm i}^{2}}{4}
    \right) \right. - \left. \ _{1}F_{1} \left(
    \frac{n+1}{2}, \frac{1}{2}, - \frac{x_{\rm i}^{2}}{4} \right) \right] .
    \label{p7}
\end{equation}
This function is flat near the center and only at large distances from the
peak it approaches the $x_{\rm i}^{-(n+3)}$ power-law applied in
\cite{hs}.

Although in the flat Universe any overdense region bounds the mass up to
infinite distance, in reality there are always neighbouring fluctuations
that also gather mass. As a way to emulate this conditions I propose a
second modification of the initial density distribution in the form of a
cut-off. One can think of two ways of estimating the cut-off scale. First,
such scale could be found as a coherence scale of the overdense region
defined by the expected overdensity (\ref{p7}) being equal to its rms
fluctuation. It turns out however, that a more stringent constraint is
induced by the presence of other peaks (see \cite{lo}). Therefore here the
cut-off will be introduced at the half inter-peak separation $x_{\rm
i,pp}/2$ for the most reasonable height of the peak, $a=3$. In the case
of $n=-1$ we have $x_{\rm i,pp}/2$=6.45. The generalized initial density
distribution with a cut-off will be modelled by
\begin{equation}    \label{p35}
    \Delta_{\rm i,cut}(x_{\rm i}) = \frac{\langle \Delta_{\rm i}(x_{\rm i})
    \rangle} {1+{\rm e}^{(x_{\rm i} - x_{\rm i,pp}/2)/w}}
\end{equation}
with the width of the filter $w=1$.

According to the SIM the subsequent shells numbered by the coordinate
$x_{\rm i}$ will slow down due to the gravitational attraction of the
peak, stop at the maximum radius and then collapse by some factor $f$ to
end up at the final radius
\begin{equation}    \label{p37}
    x = \frac{x_{\rm i} f [\Delta_{\rm i,cut}(x_{\rm i}) +1]}
    {\Delta_{\rm i,cut}(x_{\rm i})} .
\end{equation}
The simplest versions of the SIM adopt the value $f=1/2$ motivated by the
virial theorem and it will also be assumed here but a more realistic
description can be found in \cite{lo}. The final profile of the
virialized halo is then
\begin{equation}    \label{p36}
    \frac{\rho}{\rho_{\rm crit,0}} = (1 + a \sigma \varrho) (1+ z_{\rm
    i})^3 \left(\frac{x_{\rm i}}{x} \right)^2 \frac{{\rm d} x_{\rm
    i}}{{\rm d} x}
\end{equation}
where $\varrho = \xi_{R}(r)/\sigma^2$ is the correlation
coefficient.

\section{Comparison with the universal profile}

Since the measurements of halo properties from $N$-body simulations
\cite{nf} were done at the state corresponding to the present epoch, the
same condition will be applied for SIM calculations. Once the initial
redshift $z_{\rm i}$ is specified, equating the collapse time to the
present age of the Universe determines the overdensity of the presently
virializing shell which ends up at the virial radius of the halo. When we
adopt the normalization of the initial power spectrum ($\sigma_{8} = 1$)
and the conditions $a=3$ and $a \sigma=0.1$ (for the linear theory to be
valid) choosing the initial redshift $z_{\rm i}$ for a given spectral
index $n$ gives the comoving smoothing scale $R$ with which the
overdense regions are identified. The mass of the halo within the virial
radius $x_{v}$ can then also be determined
\begin{equation}    \label{p40}
    M = \frac{800 \pi}{3} \rho_{\rm crit,0} \left( \frac{x_{v}
    R}{1+z_{\rm i}} \right)^3 .
\end{equation}

\begin{figure}
\begin{center}
    \leavevmode
    \epsfxsize=8cm
    \epsfbox[96 77 372 353]{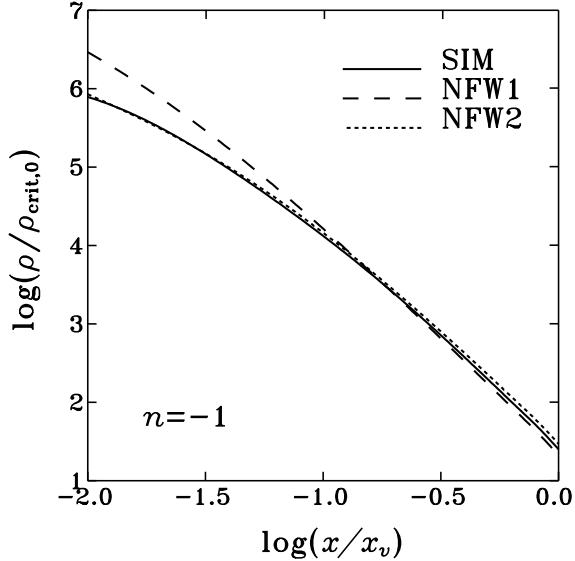}
\end{center}
    \caption{The density profiles of the halo of mass of order $3 \times
    10^{12} h^{-1} M_{\odot}$ for $n=-1$. The solid line shows the
    prediction of the SIM. The long-dashed one (NFW1)
    gives the result of the $N$-body simulations with their fitted
    concentration, while the short-dashed curve (NFW2) presents the
    formula (\ref{p20}) with concentration fitted to SIM results. }
\label{snm1a}
\end{figure}

Figure~\ref{snm1a} shows the density profile of a dark matter halo of
galactic mass. The solid line presents the prediction of the SIM obtained
from formula (\ref{p36}) for $n=-1$, $z_{\rm i} = 600$ and $R=0.188 h^{-1}$
Mpc. The final (virial) proper radius of the halo is $r_{v} = 0.231 h^{-1}$
Mpc and the mass $M=2.88 \times 10^{12} h^{-1} M_{\odot}$ which correspond
to a galactic halo. We see that the result of the SIM can be well fitted
by formula (\ref{p20}) but the SIM profile is significantly flatter than
the corresponding one from the simulations. The concentration parameters
which measure the steepness of the profile (the higher $c$ the steeper the
profile) are $c=57.1$ and $c=19.1$ respectively from the simulations and
from the SIM.

One of the main results of $N$-body simulations \cite{nf} was the
dependence of the shape of the density profiles of halos on their mass. On
the other hand, the standard prediction of the SIM
gives the same profile independently of mass. However, with
the improvements introduced above it is possible to reproduce the
dependence of the profiles on mass.

It is sometimes argued that if the density field is
smoothed with a given scale $R$ lower peaks end up as galaxies and
higher ones as clusters. This, however, would violate the hierarchical
way of structure formation since higher peaks collapse earlier. Another
argument against such assumption comes from the calculations based on the
improved SIM: the reasonable range of peak heights $a$ between 2 and 4,
which are most likely to produce halos, leads for a given smoothing scale
to the range of masses spanning only one order of magnitude, while in
$N$-body simulations halos with masses spanning few orders of magnitude
are observed. This suggests that the dependence on mass should rather be
related to the initial smoothing scale.

\begin{figure}
\begin{center}
    \leavevmode
    \epsfxsize=8cm
    \epsfbox[96 77 372 353]{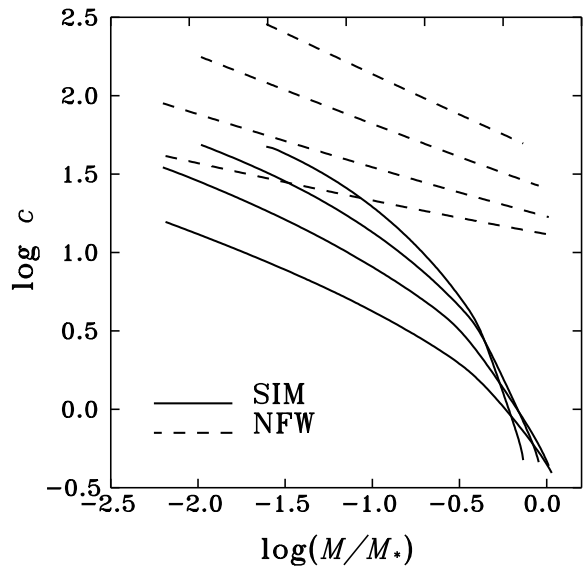}
\end{center}
    \caption{The dependence of the concentration parameter $c$ on
    mass (given in units of the present nonlinear mass $M_{\ast}$). The
    dashed curves show the results of $N$-body simulations and the solid
    lines represent the results of the SIM. In each set the curves from
    top to bottom correspond to $n=0,-0.5,-1$ and $-1.5$ respectively.}
\label{kon1}
\end{figure}

The dependence of the shape of the profiles on mass obtained with these
assumptions is shown in Figure~\ref{kon1}. The solid lines give the values
of the concentration parameter $c$ obtained by fitting the formula
(\ref{p20}) to the results of SIM for different power spectra and the
dashed lines are the corresponding values from the simulations \cite{nf}.
The overall trend of steeper profiles for smaller masses is reproduced and
the agreement between the two approaches is significantly better for
smaller masses.

The spherical infall model provides simple understanding of the dependence
of the shape of the halo on its mass: smaller halos start forming earlier
and by the present epoch their virial radii reach the cut-off scale
that accounts for the presence of the neighbouring fluctuations; more
massive halos form later and their virial radii are not affected by the
cut-off scale, their virialized regions contain only the material that
initially was quite close to the peak identified with the smoothing scale
corresponding to the mass.

\acknowledgements {This work was supported in part by the Polish State
Committee for Scientific Research grant No. 2P03D00815.}

\vfill
\end{document}